# Legal and ethical considerations regarding the use of ChatGPT in education

Fereniki Panagopoulou[1], Christina Parpoula[2], Kostas Karpouzis[3]

[1] Department of Public Administration, Panteion University of Social and Political Sciences, Athens, Greece

[2] Department of Psychology, Panteion University of Social and Political Sciences, Athens, Greece

[3] Department of Communication, Media and Culture, Panteion University of Social and Political Science, Greece

E-mail: **fereniki@panteion.gr**, **chparpoula@panteion.gr**, **kkarpou@panteion.gr**

**Abstract**

Artificial intelligence has evolved enormously over the last two decades, becoming mainstream in different scientific domains including education, where so far, it is mainly utilized to enhance administrative and intelligent tutoring systems' services and academic support. ChatGPT, an artificial intelligence-based chatbot, developed by OpenAI and released in November 2022, has rapidly gained attention from the entire international community for its impressive performance in generating comprehensive, systematic, and informative human-like responses to user input through natural language processing. Inevitably, it has also rapidly posed several challenges, opportunities, and potential issues and concerns raised regarding its use across various scientific disciplines. This paper aims to discuss the legal and ethical implications arising from this new technology, identify potential use cases, and enrich our understanding of Generative AI, such as ChatGPT, and its capabilities in education.

## 1    Introduction

A new technological tool is now available to us, under the guise of an application for compiling complex scientific answers with the assistance of artificial intelligence. But is this a blessing for learners and a curse for educators? As we all know, nothing in life is ever solely black or white, as things are usually a shade of grey. Thus, when it comes to this matter, too, attention and deliberation are required before making any aphorisms. What is clearly emerging, however, is a discernible change in the rules of the game (Mitrou, 2023: 17), as well as a valuable opportunity to provide a truly adaptive and meaningful learning experience. Therefore, this contribution aims to discuss the legal and ethical implications arising from this matter and propose ways in which the education community could use this emerging technology.

The rest of the paper is organized as follows. In Section 2, some clarifications on terminology are made. In Section 3, a brief literature review related to the use of ChatGPT in education is presented. In Section 4, a number of scenarios in which the underlying technology behind ChatGPT can improve the teaching and learning experience are discussed. In Section 5, the main legal issues that ChatGPT tool has posed are discussed in detail. In Section 6, multidisciplinary perspectives on opportunities, challenges and implications of ChatGPT for scientific research, practice and policy are presented. Finally, in Section 7, some concluding remarks are made. The bibliographic references are listed at the end of the paper in Section 8.

## 2    Some clarifications on terminology

Generative pre-trained transformer (GPT) technology is part of the family of Large Language Models that are used, inter alia, to compose/generate text by successively predicting words from other words, but without specifying the datasets it creates (Brants, et al., 2007). It is not, in fact, a new technology: it has been around for some years, but it is now being made available to the public for the first time, free of charge, and mature enough to be deployed in commercial applications and in disciplines outside natural language processing or content creation. ChatGPT is, in essence, a chatbot, to which the user enters a *prompt,* i.e., textual input which provides the context for the required response and additional instructions on writing style. ChatGPT then composes its response based on its training, the context of interaction and, more recently, information retrieved from the Web in real time. Its answers can be extensive and personalized, storing and integrating the history of the conversation, and offering users the illusion of having interacted with a natural person. According to the definition provided by the software itself in response to a relevant question, ChatGPT is an artificial intelligence program that can chat with people and answer questions. Until now, the answers have been provided without reference to the sources. Hence, it could be compared to a student who has read the course material but lacks critical thinking skills (Karpouzis, 2023).

## 3      ChatGPT in education: selective literature review

Since its public release on November 30, 2022, ChatGPT five months after its launch has already experienced a rapid growth and widespread adoption, becoming one of the most popular artificial intelligence user applications in history, so far reaching over 173 million active users. This unprecedented ChatGPT's success has posed new challenges and possibilities to a plethora of scientific domains such as finance, healthcare, medicine, material science and engineering, customer management etc. The role of this cutting-edge piece of technology in the global education field also constitutes one of the main areas of interest and contention between academics, researchers, practitioners and teachers worldwide, with a significant portion of them viewing ChatGPT as an alternative vehicle to improve and promote learning, and manage heavy workload in education as well, while others view it as a threat to integrity which opens the door to artificial intelligence-assisted cheating and/or plagiarism (Kasnecki, 2023).

GhatGPT's impact on the sector of education and lifelong learning was first systematically explored in a review by Mhlanga (2023). Mhlanga adopted the document analytical method for his research and 8 ChatGPT-related articles were finally selected to be included in his investigation in order to outline the concerns and opportunities regarding ChatGPT's use in education. According to his findings, educators expressed serious concerns and worries that students are likely to outsource their work to ChatGPT because of its ability to content creation and rapid generation of humanlike, convincing and comprehensible texts. Further, he gave emphasis on the importance of taking steps to ensure that ChatGPT is used responsibly and ethically in education, and highlighted that privacy, fairness, non-discrimination and transparency should be guaranteed. Recently, in a systematic review preprint examining 60 ChatGPT-related articles, the potential limitations and future perspectives as regards ChatGPT's use in healthcare education were investigated by Sallam (2023). The author's findings indicated that ChatGPT's benefits were expressed in 85% of the records with the most frequent being its usefulness in writing academic assignments and scientific papers, while possible risks of ChatGPT's use were cited in almost 97% of the records with the most prevalent being a plethora of ethical issues (such as bias, lack of originality, inaccurate responses etc.) and citation/referencing errors. Further, Lo (2023) reviewed 50 ChatGPT-related articles investigating how ChatGPT is utilized across various scientific fields including education. His research findings highlighted that ChatGPT is capable of revolutionizing the educational landscape if it is adopted as an instructors' assistant and as a students' virtual tutor; however, he also expressed serious concerns as regards the threats posed to academic and teaching ethics and integrity, and raised misinformation, disinformation, and mal-information issues related to ChatGPT's artificial intelligence-generated content.

It can therefore be seen that ChatGPT represents a transformational tipping point in the evolution of education and requires a more comprehensive investigation and deeper understanding of the benefits, challenges and the implications of ChatGPT-assisted learning for both educators and learners. From this perspective, it is necessary to adopt an Explainable Artificial Intelligence (XAI) approach in education, since XAI addresses four traditional moral principles; beneficence, non-maleficence, autonomy, and justice; this thereby seems to be the best way to improve trust and ethical practice in "algorithmic" educational contexts, as also discussed by Farrow (2023). Since artificial intelligence technologies have already been widely used to educational institutions serving various learning, research and practice purposes, it is of crucial importance to understand the nature of XAI in education, determine what might make it effective, and identify any ethical or practical limits in teaching and learning processes.

Undoubtedly, artificial intelligence black box technologies, such as ChatGPT, create mistrust, thus we need to bridge the gap in artificial intelligence explainability and understanding in order to comprehend artificial intelligence bias and drift, and further shape and plan the delivery of education using such technologies. An open, transparent and explainable Artificial Intelligence is the key to opening the ChatGPT's black box (Parpoula, 2023) and has the potential to improve ChatGPT's performance and provide insight into how the model is making decisions and learning. Through greater accountability and legibility, training of the involved stakeholders in ethical and legal perspectives, qualification programmes in artificial intelligence-related ethics, and greater public awareness of artificial intelligence, XAI can be a retort to the black box problem which responds with transparency to foster trust allowing users understand ChatGPT's reasoning and learning process. Since other discernible ChatGPT-related changes and implications in education have yet to emerge, continued monitoring of ChatGPT's automation, security and performance is warranted in order to ensure that ChatGPT's advantages in education are optimized, while its drawbacks are minimized.

# 4      Generative AI as an opportunity for education

The immediate response from the education community, as soon as ChatGPT was introduced, pointed to it being a risk or a threat, allowing students to plagiarize, and limiting their creativity, while reducing the individual differences between different authors (Dwivedi, 2023). However, educators and researchers also identified a number of scenarios in which Generative AI, the underlying technology behind ChatGPT, can greatly improve the teaching and learning experience. For example, educators can utilise ChatGPT to create role-playing exercises or simulate the writing style of famous authors; in this manner, the text generated can be used to attract students not interested in the mainstream teaching style, but find, for instance, contemporary music, more relatable. By adapting a generated or existing text to the style of, for example, a rap singer or a K-Pop artist, educators manage to retain the scientific integrity of their educational content, while increasing its relevance.

Another option is that of generating pros and cons with respect to a specific issue; ChatGPT has the potential to "humanize" web search, i.e., help users locate and retrieve information in the same manner as asking a fellow or colleague. A set of pros and cons can be used either as part of a more general research project or as part of a debate exercise, where students are asked to support or find weaknesses to a specific argument. Besides the actual scientific value of this generated content, a well-structured debate can be used towards improving social and soft skills, such as citizenship.

A third use which is very popular among educators is that of adapting an existing or generated text to a specific audience. In this case, educators can adapt their content with respect to scientific depth or language, making it more relevant to students with different skills and competencies. This approach is very popular with language learning (Tsatiris, 2021), and recently found its way to commercial applications, such as Duolingo. In this context, educators, either humans or an application, can select the suitable content with respect to the learning objectives of a particular module, the individual learning needs and preferences of a student, and the means of presentation and testing predicted to be more interesting for them. In this manner, students spend more time with the learning application and focus on the aspects needed for them to improve. Since ChatGPT was trained with an abundance of text from Wikipedia, books, and blog posts, its ability to generate textual content is a perfect match for language learning applications.

# 5      The emerging legal issues

This new artificial intelligence tool gives rise to several legal issues. The main ones are outlined below:

## 5.1     Issues related to plagiarism

In the case at hand, the person signing the text appears to have drafted something that is not the product of his or her intellectual property, but rather that of a third party who, in this case, is no longer a natural person, but a digital technology. Consequently, it appears that some form of cheating is taking place (Karabatzos, 2023), even though the boundaries between compositional work and the examination of sources undertaken by search engines are clearly permeable. Submitting a paper under these circumstances is against the rules of academic ethics. Indeed, it should be noted that, in accordance with Article 197(2)(b) of Law No. 4957/2022, it is a disciplinary offence "to plagiarize or conceal the direct or indirect contribution of other persons to the subject of scientific work or research". Having said that, such plagiarism can now be checked electronically and by using artificial intelligence methods, such as, for example, the Turnitin application (https://www.turnitin.com/solutions/ai-writing), although there is other software available that can 'trick' Turnitin.

In view of the above, one may rightly wonder whether what we are faced with, in a wider sense, is the art of deception. The reality of the matter is that the currently available copying methods exhaust the imagination of the examinees, who resort to all possible means for doing so, ranging from cribbing, and having other parties compose their coursework, to the use of technical means, such as mobile phones and the software in question (Koulouri 2023). And, if we were to go a little deeper, we might end up concluding that our entire culture is a copycat, the requirement being that it be a good one.

## 5.2     Issues concerning copyright

Another question that arises in this context is who should be considered the final work's author. The possible answers can be summarized as follows (Panagopoulou-Koutnatzi, 2023; Chiou, 2022, 2021, 2017):

   a) The work is the property of the creator of the artificially intelligent software. This position is countered by the fact that the application of the ideas of a creation does not constitute a derivative work, as ideas do not fall within the scope of copyright protection under Art. 2 of Law 2121/1993.
b) The work that is generated belongs to the creator of the artificially intelligent software infrastructure as intellectual property, but not as copyright. In this sense, the work created may be considered intellectual property belonging to the creator of the artificial intelligence, e.g., as an invention, but not his or her copyright. In this sense, the intellectual creation itself might belong to the user of the creative artificial intelligence and not to the creator of the artificial intelligence. Since work suffices as a criterion for the acquisition of intellectual creation, following Art. 1(3) of Directive 91/250/EEC and Article 6 of Directive 2006/116/EEC, the secondary achievements would be the property of the user of the original software, since it was the user who put the device in question into operation to produce them (Christodoulou, 2019: 122-123).
c) The produced work comprises the joint creation of the creator and the user of the creative artificial intelligence. This approach takes into consideration the fact that the final, jointly created work is the co-creation of both parties (Christodoulou, 2019: 122-123).
d) The created work becomes a free good, which is now in the public domain, since machines cannot create intellectual works (Christodoulou, 2018: 54, note 119). In this case, we are dealing with the so-called zero-sum solution. In this context, the secondary creation is not the product of a natural person and, as such, it is not a work, but rather a free good belonging to the public domain. Even so, ownership of the product of the artificial intelligence will be acquired by its owner through the processing of material belonging to a third party or as fructus (Panagopoulou-Koutnatzi, 2023: 54, note 122). This solution is founded on the argument that what we have at hand is not a human creation that is tangible and original. At the same time, this approach has the disadvantage of lacking any motivation for the manufacturer of the artificial intelligence (Igglezakis, 2022: 214). Moreover, it is maintained that free distribution is inconsistent with the Berne Convention, from which the law of copyright derives, and which also establishes the principle of the author.
e) The work generated is the product of the creative software, meaning that the artificial intelligence becomes a creator from the position of the creation (Zekos, 2022: 80). In this way, the legal personality of the artificial intelligence device is acknowledged either by analogy or by virtue of legislation. The European Parliament (EP) resolution of 16 February 2017 on "Civil Law Rules on Robotics", which was rejected by the EP in October 2020, moves towards the direction of establishing a special legal framework for robotics in the long term. The aim is to consider more sophisticated, autonomous robots as electronic persons, with the obligation to rectify any potential damage caused. A further aim is to apply (legal) electronic personhood in cases where robots make autonomous decisions or otherwise interact independently with other persons (EP resolution on "Civil Law Rules on Robotics" 2017). This position entails the risk of limiting responsibility for potential damage to the benefit of the devices' manufacturers and has not been endorsed by the European legal order (Panagopoulou-Koutnatzi, 2023: 123). In the Thaler v. Comptroller judgment issued by the England and Wales Court of Appeal (EWCA) in 2021, it was held that a machine cannot be an inventor within the meaning of the law, as a machine is not a natural person (EWCA 2021). In China, by contrast, it has been ruled that an article generated by a robot is protected by copyright (Sawers, 2020).
f) In view of the weaknesses entailed in the above positions, the solution of unjust enrichment is proposed, under the lens of civil law, pursuant to the provisions of Art. 904 et seq. of the Civil Code. In this case, ownership is acquired as unjust enrichment deriving from a lawful cause (by virtue of a contract) or even without lawful cause.

None of the above solutions can be said to wear the crown of absolute rightness, and the answer to the question of copyright must be given based on the particular facts of each case. It is anticipated, however, that artificial intelligence will necessitate the transformation of the law on copyright, which may end up having to attribute rights to non-human creators (Chiou, et al., 2016).

## 5.3    Issues of legal responsibility

If the drafted document is adopted, it is only reasonable that the issue of liability should arise. Who will be responsible if the created work contains false statements? Quite evidently, this is a question that does not lend itself to an obvious answer.

The first position is that responsibility must be borne, but also managed, by the manufacturers of artificial intelligence products. This could be achieved by establishing a rebuttable presumption providing that, in case of doubt, manufacturers shall be deemed responsible. This would strengthen responsibility and foresight on their part. To this end, it would seem appropriate that an impact assessment should be required before activating any artificial intelligence application (see, for example, Article 5 of Law No. 4961/2022). This model of responsibility seems to be largely adopted by the Draft Regulation on Artificial Intelligence, which assigns a great deal of responsibility to the software designer and emphasizes the importance of forethought at the design stage. Furthermore, AI companies are under obligation to exercise rigorous after-sales control over their products and to conduct continuous upgrades that will prevent unforeseen impacts (Kowert, 2017: 203). Therefore, AI companies must devise ways to prevent the misuse of their products in order to protect themselves but also to avoid depriving society of the great benefits that they will offer. They must take appropriate measures to minimize the risks that may arise while, in doing so, they will also reduce their potential responsibility and ensure that their products are suitable for the society in which we live (Kowert, 2017: 203). No one should develop artificial intelligence systems without having a sense of responsibility for them, even if they are autonomous machine learning systems, since responsibility can now also be introduced as information (Winfield & Jirotka, 2018). Strict liability on the part of the creator should play a key role in terms of compensating for damage caused by defective products and their components, whether they come in tangible or digital form (European Commission, 2019: 8).

The second position lies in the view that responsibility should be attributed to the user of the technology that the intelligence involves, i.e. the researcher who applies the technology. This does not, however, resolve all the issues raised by artificial intelligence: still, it remains a prima facie honest solution as far as the researcher seeking a proposal for the problem at hand is concerned. Rather than embracing the proposal without question, the user of the program in question ought to check that the proposal is fully adapted to the facts of the case under consideration and consider the possible scenario that the algorithm may be biased. In this direction, we could adopt the rebuttable presumption that human judgement prevails over the algorithm's decision in case of doubt (cf: Article 22(3) GDPR). Of course, the risk of the user being carried away by the proposal and being led to misguided reasoning when it comes to the final text should not be underestimated.

The third position is based on the sharing of responsibility between the manufacturer or developer of the artificial intelligence technology and its user. Each of them will be responsible for his or her share of responsibility: the developer for the manufacturing defect and the user for the failure in handling it or for not taking into account the facts in the case under consideration. Even though this system of responsibility seems appealing and appears to be the most prevalent one, it also comes with its own problems and controversies. Attribution of responsibility may in many cases be rendered an issue that is difficult to solve and prove. If there are two or more actors, in particular (a) the person who primarily makes the decision on the use of the relevant technology and benefits from it (frontend operator); and (b) the person who continuously determines the characteristics of the relevant technology and provides substantial and ongoing support to the backend (backend operator), objective responsibility should rest with the person who has greater control over the risks of the operation (European Commission, 2019: 8).

The fourth position involves the attribution of responsibility to the technology itself. But, if the technology is to be held responsible, it must first be granted legal personality (Papakonstantinou & De Hert, 2020). In 2015, the EP adopted a resolution inviting the Commission to consider the possibility of creating a special legal status for robots in the long term (Zornoza, et al., 2017; Borenstein & Arkin, 2016; Deng, 2015; Lin & Bekey, 2015; Veruggio & Abney, 2011). The aim of this would be to have at least the most sophisticated, autonomous robots acquire the status of electronic persons responsible for rectifying any damage they may cause, and possibly also the recognition of electronic personhood in cases where robots make autonomous decisions or otherwise interact with third parties independently.

This solution was rejected in October 2020 by the EP committee (2016) which adopted three resolutions on the ethical and legal aspects of artificial intelligence software systems, namely a) Resolution 2020/2012 (INL) on a framework of ethical aspects of artificial intelligence, robotics and related technologies; b) Resolution 2020/2014(INL) on a civil liability regime for artificial intelligence; and c) Resolution 2020/2015(INI) on intellectual property rights for the development of artificial intelligence technologies. All three resolutions acknowledge that artificial intelligence will have significant benefits across various sectors (businesses, the labour market, public transport, and the health sector).

Even so, as pointed out in the resolution on the ethical aspects of artificial intelligence, there do exist concerns that the current legal framework of the European Union, including consumer law, labour law and social acquis, data protection legislation, product safety and market surveillance legislation, as well as anti-discrimination legislation, may no longer be adequate to effectively address the risks posed by artificial intelligence, robotics, and related technologies. All three resolutions are unequivocal in not granting legal personality to artificial intelligence software systems. Consequently, tempting as it may be, it appears that this solution will not be adopted in the near future, although it is not ruled out for a little later when the concept of digital personality will have matured.

In any event, the proposal for a Directive on adapting non-contractual civil liability rules to artificial intelligence is a step in the right direction, which will create a rebuttable "presumption of causation" to ease the burden of proof placed on victims, who must prove the damage caused by an artificial intelligence system. Additionally, it provides national courts with the power to order the disclosure of evidence relating to high-risk artificial intelligence systems that are suspected to have caused damage.

### 5.4 Issues regarding the freedom of expression

A question that arises is whether the software is covered under the freedom of expression (Massaro & Norton, 2016: 1169). More specifically, does the machine have the discretion to characterize someone as a famous or obscure professor or a notable scholar? If we, as ordinary citizens, ask questions to a journalist and the journalist answers them, it is indisputable that the journalist is covered by the constitutionally guaranteed freedom of speech. Similarly, when we submit a question to the software it must decide, at that moment, which "answers" it should give us and in what order. If those answers are regarded as an expression of the software, then any governmental attempt to regulate the technology should be regarded as censorship (Wu, 2013: 161). To the extent that the developer of the software incorporates his or her opinion and attempts to influence the public, freedom of speech is assumed to apply in this case (Wu, 2013: 1533). There are considerable concerns about the misinformation (Tsakarestou, 2023) of citizens through the re-dissemination of false news.

### 5.5 Issues pertaining to the protection of personal data

The use of large language models in education raises concerns over privacy and data security, as learner data are often sensitive (special categories). The indiscriminate collection and processing of our personal data resulting from the operation of artificial intelligence gives rise to intense questions about the compatibility of the technology with the right to personal data protection and that of informational self-determination. A large amount of data used in artificial intelligence constitutes personal data (Igglezakis, 2022: 175) and much of it falls into special categories.

By way of explanation, the operation of artificial intelligence requires the collection and processing of large data sets that are difficult to put under the control of the data subject. As the algorithm often outperforms its creator, due to the latter's inability to comprehend the way in which it operates, it is not always possible to inform the data subject on how the algorithm works and, by implication, on the data being collected and its wider processing. As a result of this, the principle of transparency is not adhered to.

At the same time, inaccuracies concerning persons give rise to questions concerning the violation of the principle of data accuracy. Other usual risks lie in the unauthorized access to learners' data and the use of such data for purposes beyond those related to education. In this respect, the Italian Data Protection Authority has ordered the temporary restriction of the processing of Italian users' data against OpenAI, the American company that developed and manages the ChatGPT platform. In parallel with this, the Authority has also initiated a related investigation (GPDP, 2023).

The Italian Data Protection Authority has highlighted the lack of information for users and all interested parties whose data are being collected by OpenAI, and especially the absence of a legal basis justifying the mass collection and storage of personal data for the purposes of "training" the algorithms underpinning the operation of the platform. Despite the fact that, according to the terms published by OpenAI, the service is intended for users aged 13 and over, the Italian Authority noted that the lack of any mechanism for verifying the age of users exposes minors to responses that are wholly inappropriate for their level of development and understanding. This clearly raises the issue of the responsibility of the controller to take appropriate technical and organizational measures to prevent children from having access to this type of software (Panagopoulou-Koutnatzi, 2017: 51). OpenAI, which does not have an establishment in the European Union but has appointed a representative in the European Economic Area, must notify the measures it has taken to implement Garante's request within 20 days, subject to a fine of up to EUR 20 million or up to 4% of its global annual turnover.

### 5.6 Risks posed against the liberal character of the democratic political system

The imposition of a "dictatorship" of the average in science, the provision of specific, premeditated, tested knowledge, poses the danger of undermining the liberal character of our constitution, in the sense of imposing the average in science and in our thinking in general (Foundethaki, 2023), of establishing a common understanding of things, but also of spreading misleading or false news. It is reasonable to wonder whether we should intervene legislatively to preserve the core of liberal democracy. This question is a multifaceted one, as the establishment of a particular perception leads, in turn, to the formulation of a specific electoral preference and, in this instance, to the indirect manipulation of voting.

# 6 What would be the appropriate response on the part of the scientific community?

It is true that this new technology has been a source of great concern in the educational community. Have we arrived at the death of the author (Barther,1968) or the reader (Vamvakas, 2023)? There is talk of the depletion of the knowledge ecosystem, and this is because the development of knowledge has, as its starting point, the matters that the scholar is immersed in, such as published scientific papers and books that build on previous knowledge (Spinelli, 2023). Education stems from what has educated the educator who, after the "ordeal" of intellectual pursuit, can pass on the knowledge acquired to his or her students. The "un-educated" response to any question posed seems to deprive learners of the necessary interactions with the knowledge ecosystem (Spinelli, 2023). The automatic answer deprives us of the journey of knowledge. Fears are expressed about the impairment of original intellectual creativity and critical thinking (Karabatzos & Skevi, 2023). Could it be said that we have come to the end of the age of conventional writing skills and education at large? Is this an intellectual revolution (Parpoula, 2023), a hoax, or a commonplace of evil (Chomsky, 2023)? It would be best if we did not hasten to rash conclusions.

The question, however, remains a vexing one: how should we approach the issue of software? A solution that has been put forward is to discard technology, essentially turning the clock back to the 20th century, and have students take their exams with pen and paper, without the use of electronic devices connected to the internet (Villasenor, 2023). For instance, the University of California in Los Angeles is looking into the possibility of making it a violation of its honour code of ethics to use ChatGPT to take exams or write papers (Villasenor, 2023). Likewise, in Germany, the University of Tübingen has decided to restrict the use of this software for students and researchers (Universität Tübingen, 2023).

The reasons for this hesitation are not problem-free. Learners may rely excessively on this model, but information that is generated effortlessly could adversely affect critical thinking and problem-solving skills. This is because this model simplifies the acquisition of answers or information—and this is something that could reinforce the learners' laziness, as well as limit their interest in conducting their own research to reach their own conclusions or solutions (Enkelejda, et al., 2023).

Nevertheless, banning technology would be a technophobic approach if it were to be taken indiscriminately. The voices of those who urge their students to use ChatGPT in their written assignments also entail certain concerns. Instead of banning learners from using artificial intelligence writing tools that can save them time and effort, we should teach them how to use them ethically and productively (Villasenor, 2023), to enable them to comprehend the complex matter of diversity of sources and the educational process in general. In this sense, choosing to exercise strict control over the system would be preferable.

To remain competitive throughout their careers, learners need to be trained on how to prompt an artificial intelligence writing tool to generate a meaningful output and assess its quality, accuracy, and originality (Villasenor, 2023). For this reason, the software should be a tool that will support each related course (Pedis & Karpouzis, 2023). Learners must be taught how to write well-structured, coherent essays incorporating a combination of artificial intelligence-generated text, along with traditional writing. As professionals, they need to learn how to work productively with artificial intelligence systems, utilizing them to complement and enhance human creativity with the extraordinary potential that they promise to bring to the table (Villasenor, 2023).

In addition to pedagogical reasons for approaching ChatGPT as an opportunity, and not as a threat, there are also practical reasons for doing so: apart from concerns regarding academic freedom (Parpoula, 2023), it is also utopic to effectively prohibit access to this technology (Villasenor, 2023). This software is freely available, and learners cannot be monitored in the free space of their private life (Karabatzos & Skevi, 2023). The reasoning for supporting a total ban clearly does not solve the problem at hand (Floros ,2023).

The imposition of a total ban on the use of ChatGPT would also inevitably result in the injustice of false positive and false negative results in the course of monitoring the use of the software. Some learners who use ChatGPT despite the ban could, either by chance or thanks to the rather thorough processing of the text that is generated by artificial intelligence, avoid having their text flagged out as being assisted by it. In a worse scenario, some learners could be falsely accused of using ChatGPT, causing immense anxiety and leading to penalties for a slip they did not actually make (Villasenor, 2023).

It would be presumptuous to ignore an application that offers personalized instruction and feedback to learners based on their individual learning needs and their progress. For instance, the application could provide personalized math instruction to learners, resulting in improved learning outcomes (Baidoo-Anu & Owusu Ansah, 2023). The application could assist in the development of reading and writing skills (for example, by suggesting syntactical and grammatical corrections), as well as in the development of writing and critical thinking skills (Enkelejda et al., 2023). These models can also be used in the creation of questions and as prompts that will encourage learners to think critically about what they read and write, and to analyze and interpret the information they are presented with (Enkelejda et al., 2023). This does not mean that the software can replace teachers, but rather that it can assist them (Baidoo-Anu & Owusu Ansah, 2023).

Furthermore, we should also not overlook the potential of the software to empower learners with disabilities. Language models can be used to develop inclusive learning strategies with adequate support for tasks such as adaptive writing, translation, and the identification of important content in various formats (Enkelejda et al., 2023). Still, it is important to note that the use of large language models should be supplemented with the assistance of professionals, such as speech therapists, teachers, and other specialists who will be able to adapt the technology to the specific needs of the learner's disabilities (Enkelejda et al., 2023).

Writing a good essay from scratch requires careful, and often painstaking thinking about its structure, flow, and delivery. This can be developed as a skill in early education classes. Learning to write without the use of artificial intelligence does, indeed, promote focused, disciplined thinking. But learning how to successfully combine conventional writing with the support of artificial intelligence to create really good essays also requires these skills (Villasenor, 2023). It is like turning our backs to the future, having an application at our disposal and saying that it is forbidden to use it. By the same reasoning, we could prohibit finding sources through search engines and only allow searching in conventional libraries. Is that what we wish to do? It would appear better to allow the creative use of artificial intelligence, with a view to finding ways of combining it with traditional education.

*How could this be achieved?*

1. Educators ought not to look down on the issues of new technologies, but to be trained and comprehend, in this respect, what ChatGPT is and how it works (Karabatzos & Skevi, 2023), as well as explore technological applications through the lens of academic integrity (Parpoula, 2023).
2. Learners must be prepared for a future in which artificial intelligence will be just another technological tool.
3. Learners should be solely and exclusively responsible for the texts that they hand in under their names. If they contain inaccuracies, they will be responsible for finding the truth. If their structure is problematic, they will also have the responsibility of their signature. If the text is stylistically or logically inconsistent, it will be their responsibility. If there is partial plagiarism, they will also be legally responsible for it (Villasenor, 2023). They will, thus, be responsible for checking and evaluating their sources (Karabatzos & Skevi, 2023) and, above all, they must add references to their text, which is something that the software does not provide at this time.
4. In this sense, learners should be encouraged to be responsible, informed users of artificial intelligence technologies that will play an extremely important role during their careers. This responsibility entails an obligation to report on the use of the software in the text.
5. Educators must shift their focus and place more importance on how they teach, adopting a different approach in their teaching method.
6. Without this meaning that we will have to renounce or ban technology, we could think of some alternative, complementary ways of studying and testing, such as:
    a) Cultivating conventional writing starts in the lower grades so students are not cut off from this very useful skill.
    b) The performance of learners could be (co-)assessed through conventional examinations that will not involve the use of mobile phones.
    c) It is recommended that learners should be tested orally on critical questions concerning their coursework (Lakassas, 2023).
    d) It would be better if assessment questions were to require more critical thinking to make it difficult for machines to compose them. We have a duty to educate people so that we have citizens and scientists engaged in critical thinking (Pitta, 2023) and do away with the "dictatorship" of the average.
    e) Examinations could be conducted remotely, online, using artificial intelligence systems that monitor all suspicious actions on the part of examinees, including the use of software. In fact, this is how the examinations for selecting senior civil servants in public administration were conducted. This type of examination requires a prior study of technology's impact on the

examinees' rights.
   f) Examiners are advised to submit the assessment questions to the software to enable them to be aware of the answers it generates, even though the software produces different answers each time.

# 7      In lieu of an epilogue

As is the case with every new technology, ChatGPT also poses a new challenge. It is up to us not to demonize it and, instead, proceed to the ethical use of this technology by adhering to the fundamental principles of logic, moral philosophy, and aesthetics. We ought to make use of the countless advantages of technology, assessing and managing any risks it may entail (Panagopoulou-Koutnatzi, 2023; Chiou, 2022, 2021, 2017). We have the ability to control technology with the help of technology itself, and the transparency and comprehensibility of technology will be crucial in this respect.

**Acknowledgements**

This paper is a revised version of a contribution titled "ChatGPT: Renouncing the new technology or cultivating its ethical use?", published in Syntagma Watch, 7.3.2023. Available online:
https://www.syntagmawatch.gr/trending-issues/chatgpt-apokhryjh-ths-neas-texnologias-h-kalliergeia-hthikis-xrhsews-ths/.

Fereniki Panagopoulou has prepared Sections 1, 2, 5, 6, and 7. Christina Parpoula has prepared Section 3. Kostas Karpouzis has prepared Section 4. All authors contributed to Abstract and References sections, and paper editing as well.

**Conflicts of Interest**
The authors declare no conflict of interest.